\documentclass[aps,prl,showpacs,twocolumn]{revtex4}

\usepackage{bm}
\usepackage{graphicx}
\usepackage{color}
\usepackage{afterpage}

\begin{document}

\title{%
Numerical Detection of the Ergodicity Breaking in a Lattice Glass Model
}

\author{Munetaka Sasaki}
\email{msasaki@camp.apph.tohoku.ac.jp}
\affiliation{Department of Applied Physics, Tohoku University, Sendai 980-8579, Japan}
\author{Koji Hukushima}
\affiliation{Department of Basic Science, The University of Tokyo, Tokyo 153-8902, Japan}

\date{\today}
\begin{abstract}
We directly detect the ergodicity breaking in a lattice glass model 
by a numerical simulation. The obtained results nicely agree with those by the cavity method 
that the model on a regular random graph exhibits a dynamical transition 
with the ergodicity breaking at an occupation density. 
The present method invented for a numerical detection of 
the ergodicity breaking is applicable to glassy systems in finite dimensions. 
\end{abstract}

\maketitle

Glass is a jammed state that particles can not move freely because they are densely packed. 
As the occupation density of particles increases, the system explores the phase space more and more slowly, 
and it eventually stop exploring the whole phase space. This impossibility for the system 
to explore the whole phase space is called ergodicity breaking. Although it is highly controversial 
whether real glasses in a finite dimension exhibit a dynamical glass transition with the ergodicity 
breaking or not, several theories of the glass predict the existence of the ergodicity breaking. 
The so-called random-first-order-transition 
theory~\cite{KirkpatrickThirumalai87,KirkpatrickWolynes87,KirkpatrickThirumalaiWolynes89} 
is one of them. This theory predicts that a single huge liquid state is broken up into exponentially 
large number of metastable amorphous solid states at a certain density (or temperature) so that the structural 
entropy or the complexity becomes finite. 
The (free-)energy barriers among the states become infinite in the thermodynamic limit and the ergodicity 
is broken there. Detailed analyses in statistical mechanics 
with the cavity method~\cite{MezardParisi01,MezardMontanariBook} 
indicate that the Biroli-M{\'e}zard (BM) lattice glass model~\cite{BiroliMezard01} 
defined on a regular random graph exhibits the ergodicity breaking~\cite{Rivoire04,Krzakala08} 
in accordance with the random-first-order-transition scenario. A series of studies with the cavity method 
also indicate that many constraint-satisfaction problems exhibit an analogous clustering transition 
of solutions (see~\cite{MezardMontanariBook} and references therein). 
However, it is still an open problem whether there exists the ergodicity breaking in these systems 
because the cavity method is based on several non-trivial assumptions 
(see chapter 19 of Ref.~\cite{MezardMontanariBook})
and it does not directly observe the structure of the phase space. Therefore, it is desirable 
to verify the existence of the ergodicity breaking by a numerical simulation to check the validity 
of the previous studies. Furthermore, to study the ergodicity breaking in finite-dimensional systems, 
a proper numerical method is indispensable because analytical methods such as the cavity method 
are not available there.

In the present study, we investigate the BM lattice glass model on a regular random graph 
by Monte-Carlo (MC) simulation with the aim of a direct detect of the ergodicity breaking. 
By properly designing the method and adopting a list-referring MC method~\cite{SasakiHukushima13} 
and the Wang-Landau method~\cite{WangLandau01A, WangLandau01B} 
for an efficient exploration of the phase space, we succeeded in directly detecting 
the ergodicity breaking. Our results nicely agree with those by the cavity method 
that the model exhibits a dynamical transition with the ergodicity breaking at an occupation 
density $\rho_{\rm d}=0.5708$.

{\it Model.---}
The BM model is a kind of lattice glass models. A binary variable $\sigma_i$ is defined on each site. 
The variable $\sigma_i$ denotes whether a site $i$ is occupied by a particle ($\sigma_i=1$) 
or not $(\sigma_i=0)$. In this study, we consider the BM model defined on a regular random graph. 
Each site is connected with $k$ neighbouring sites which are chosen randomly from all of the sites. 
A particle configuration $\{\sigma_i\}$ is restricted by hard constraints that neighbouring 
occupied sites of each particle should be less than or equal to $l$. The BM model is 
characterized by the two integers $k$ and $l$, which satisfy the inequality $k>l$. 
The probability distribution of the BM model for a particle configuration $\{\sigma_i\}$ is given as
\begin{equation}
P\{\sigma_i\}=Z^{-1}C\{\sigma_i\}W\{\sigma_i\}. 
\label{eqn:distribution}
\end{equation}
In this equation, $Z$ is the partition function defined by 
$Z\equiv {\rm Tr}_{\{\sigma_i\}} C\{\sigma_i\}W\{\sigma_i\}$
and $C\{\sigma_i\}$ is an indicator function which is 
one if $\{\sigma_i \}$ satisfies all of the constraint conditions or zero otherwise. 
$W\{\sigma_i\}$ is a weight of the particle configuration $\{\sigma_i\}$. 
For example, in the case of the grand-canonical ensemble, 
$W\{\sigma_i\}$ is given as $W\{\sigma_i\}=\exp\left[\mu N\{ \sigma_i \}\right]$, 
where $\mu$ is a chemical potential, $N\{\sigma_i\} \equiv \sum_{i=1}^{N_{\rm site}} \sigma_i$, 
and $N_{\rm site}$ is the number of sites. 

\begin{figure}[t]
\begin{center}
\includegraphics[width=\columnwidth]{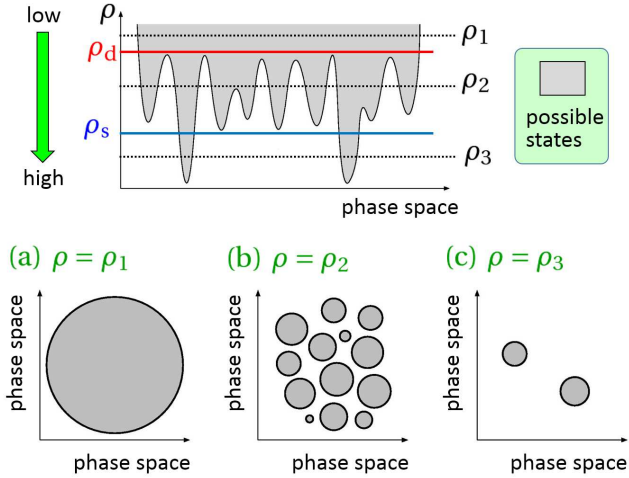}
\end{center}
\caption{(Color online) A schematic illustration of the phase space structure 
of the BM model indicated by the previous studies~\cite{Rivoire04,Krzakala08}. 
The gray region denotes possible states (particle configurations) 
which satisfy all of the constraint conditions. The vertical and horizontal  axes of the top figure are 
the occupation density $\rho$ and the phase space, respectively. 
The three bottom figures are cross sections of the top figure at (a) $\rho=\rho_1$, (b) $\rho=\rho_2$, 
and (c) $\rho=\rho_3$, respectively, where $\rho_1<\rho_{\rm d}<\rho_2<\rho_{\rm s}<\rho_3$. 
One huge cluster is divided into numerous clusters 
at a dynamical transition density $\rho_{\rm d}$, and the number of clusters becomes of order one 
at a static transition density $\rho_{\rm s}$.
}
\label{fig:ErgodicityBreaking}
\end{figure}

\begin{figure}[t]
\begin{center}
\includegraphics[width=\columnwidth]{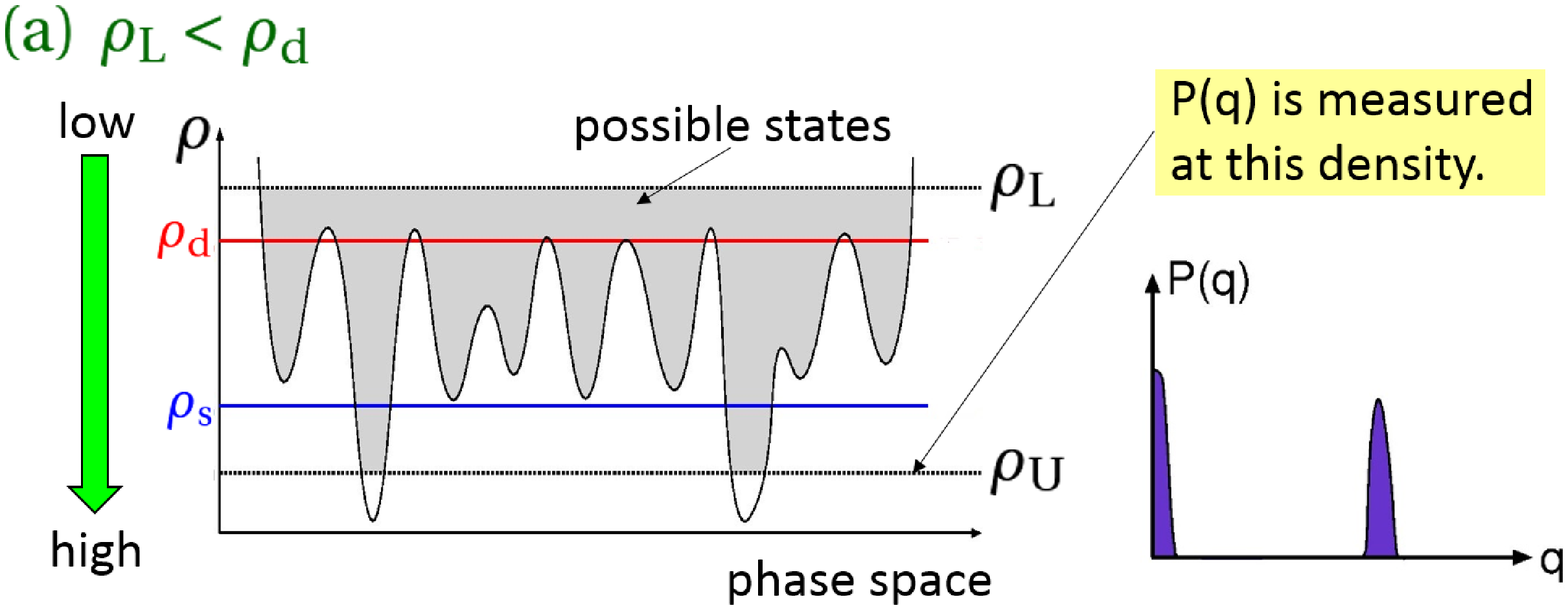}\vspace{2mm}\\
\includegraphics[width=\columnwidth]{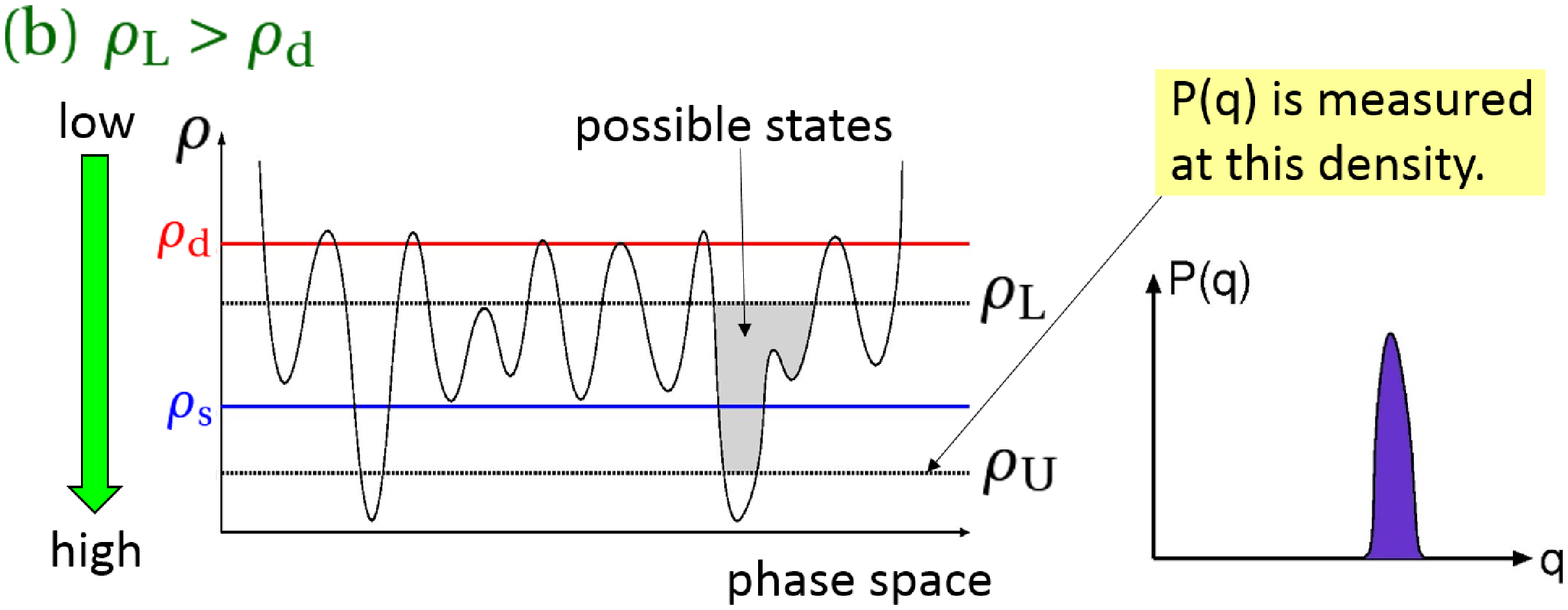}
\end{center}
\caption{(Color online) A schematic illustration of a simulation to detect the ergodicity breaking. 
The system can explore the gray regions in the left two panels under the restriction 
$\rho_{\rm L}\le \rho \le \rho_{\rm U}$. The simulation is performed for two replicas 
to measure the probability distribution of the overlap $P(q)$. The overlap $q$ is measured 
at $\rho_{\rm U}$. Two replicas have the same initial particle configuration. 
When (a)~$\rho_{\rm L}<\rho_{\rm d}$, the system can explore the whole phase space when 
it reaches low density region. Therefore, the two replicas can be in either the same valley or different ones. 
In contrast, when (b)~$\rho_{\rm L} > \rho_{\rm d}$, two replicas can never escape from the common 
initial valley. The right two panels show expected $P(q)$'s in the two cases. 
}
\label{fig:Expected_Behavior}
\end{figure}

In the present study, we will focus on the BM model on a regular random graph with $k=3$ and $l=1$. 
The phase space structure of the model indicated by detailed analyses based on 
the cavity method~\cite{Rivoire04,Krzakala08} 
is briefly illustrated in Fig.~\ref{fig:ErgodicityBreaking}. 
The model exhibits a dynamical transition at an occupation density $\rho_{\rm d}=0.5708$. 
The phase space is divided into numerous clusters at this density and the ergodicity is broken there. 
However, no static anomaly is detected at $\rho_{\rm d}$. 
As the density further increases, the number of clusters decreases and it finally becomes of 
order one at an occupation density $\rho_{\rm s}=0.5725$. A static transition 
with a one-step replica symmetry breaking occurs at this density. 
The purpose of the present study is to directly detect the dynamical transition 
with the ergodicity breaking by a numerical simulation.

{\it Basic strategy.---}
Figure~\ref{fig:Expected_Behavior} shows a schematic illustration of a simulation 
to detect the ergodicity breaking. In the present study, we perform a 
simulation  in which the occupation density $\rho$ is restricted to be 
$\rho_{\rm L} \le \rho \le \rho_{\rm U}$. Therefore, the system can only 
explore the gray region in the figures. The upper occupation density 
$\rho_{\rm U}$ is fixed to a certain value above the static transition density 
$\rho_{\rm s}=0.5725$ evaluated by the cavity method, while the lower occupation density 
$\rho_{\rm L}$ is changed across $\rho_{\rm d}$. 

To judge whether an ergodicity breaking is induced by the change in $\rho_{\rm L}$ or not, 
we measure the probability distribution of the overlap 
$P(q)$ between two states. We therefore perform simulations for two replicas. 
The two replicas have the same initial state and their particle configurations are updated 
independently with different random number sequences. The overlap $q$ is calculated from 
two states at the upper occupation density $\rho_{\rm U}$. 
In this measurement, we can expect that $P(q)$ exhibits the following $\rho_{\rm L}$ dependence: 
When (a)~$\rho_{\rm L}<\rho_{\rm d}$, the system can explore the whole phase space 
when it reaches low density region (see the top left panel in Fig.~\ref{fig:Expected_Behavior}). 
Therefore, two replicas can be in either the same valley or different ones. 
As a result, $P(q)$ has two peaks around $0$ and $1$ (see the top right panel). 
In contrast, when (b)~$\rho_{\rm L} > \rho_{\rm d}$, $P(q)$ has a single peak around $1$ 
because two replicas can never escape from the common initial valley.

{\it Devices for an efficient exploration of the phase space.---}
Following the protocol mentioned above, there is a possibility to misjudge the occurrence 
of the ergodicity breaking because of insufficient exploration of the phase space 
in numerical simulations. 
To reduce the possibility, the phase space should be explored as efficiently as possible. 
We therefore devised the simulation method in the following two points: Firstly, we used 
an efficient list-referring MC (LRMC) method~\cite{SasakiHukushima13}. 
The basic idea of the LRMC method is similar to that of the $N$-fold way method~\cite{BortzKalosLebowitz75}. 
By using a list of sites into which we can insert a particle, we avoid trying a useless transition 
which is forbidden by the constraint conditions. It is demonstrated in Ref.~\onlinecite{SasakiHukushima13} 
that the relaxation time of the LRMC method is about $10^3$ times shorter than that of 
the standard MC method in some cases. Secondly, we measured the density of states (DOS) 
and utilized it for an efficient measurement of $P(q)$. To be specific, when we measure $P(q)$, 
we perform a MC simulation in which the weight $W\{\sigma_i\}$ in Eq.~(\ref{eqn:distribution}) is given as
\begin{equation}
W\{\sigma_i\}=\frac{1}{\Omega(N\{\sigma_i\})},
\label{eqn:Weight_WL}
\end{equation}
where $\Omega$ is the DOS defined by
\begin{equation}
\Omega(N') \equiv {\rm Tr}_{\{ \sigma_i \}} \delta_{N',N\{ \sigma_ i \}} C\{ \sigma_i \}.
\label{eqn:Def_DOS}
\end{equation}
We adopted this weight so that the probability $P(N')$ of the number of particles being $N'$, 
which is defined by 
\begin{equation}
P(N') \equiv {\rm Tr}_{\{ \sigma_i \}} \delta_{N',N\{ \sigma_ i \}} P\{\sigma_i\},
\end{equation}
does not depend on $N'$. By visiting all of densities between $\rho_{\rm L}$ and $\rho_{\rm U}$ 
with an equal probability, the system can explore the phase space efficiently. 
The DOS $\Omega(N')$ is numerically calculated by the Wang-Landau 
method~\cite{WangLandau01A,WangLandau01B}, 
which is one of the standard methods for the DOS calculation. 
As demonstrated in Ref.~\onlinecite{SasakiHukushima13}, the performance 
of the Wang-Landau method is significantly improved with the combined use of the LRMC method.

{\it Procedures of the simulation.---}
In the present simulation, we first perform a simulated annealing MC 
in the grand-canonical ensemble to find a particle configuration with the occupation 
density $\rho_{\rm U}$. In this calculation, the chemical potential $\mu$ is gradually 
increased from $0$ to $12.5$ by $0.1$, and a particle configuration with a maximum 
occupation density is recorded. Because the maximum occupation density obtained 
by simulations is larger than $\rho_{\rm U}$ in almost all cases, we can create a particle 
configuration with the density $\rho_{\rm U}$ by deleting a few particles at random. 
This particle configuration is used as an initial state in both the subsequent DOS calculation 
and the $P(q)$ measurement. We next calculate the DOS $\Omega(N')$ 
within $\rho_{\rm L}\le \rho \le \rho_{\rm U}$ by the Wang-Landau method. 
The DOS is calculated for each sample of a random graph.  

After these two steps, we perform a simulation for the $P(q)$ measurement with the weight 
Eq.~(\ref{eqn:Weight_WL}). To measure $P(q)$ at $\rho_{\rm U}$, we update the 
particle configurations of the two replicas in a sequential manner: We first update 
the particle configuration of the first replica until the following two conditions are satisfied:
\begin{itemize}
\item The occupation density is $\rho_{\rm U}$. 
\item The elapsed time from the previous measurement is larger than one MC step. 
\end{itemize}
The second condition is imposed to avoid a successive measurement. We then update the particle 
configuration of the second replica until the two conditions are satisfied. After that, we calculate 
the overlap $q$ from the particle configurations of the two replicas 
$\{ \sigma_i^{(1)} \}$ and $\{ \sigma_i^{(2)} \}$, where $q$ is defined by
\begin{equation}
q\equiv \frac{1}{C_{\rm N}}\sum_i (\sigma_i^{(1)}-\rho_{\rm U}) (\sigma_i^{(2)}-\rho_{\rm U}),
\end{equation}
with $C_{\rm N}\equiv N_{\rm site}\rho_{\rm U}(1-\rho_{\rm U})$ being a normalization factor. 
The value of $q$ is close to one if the two particle configurations are similar, 
and it is close to zero if there is no correlation between them. 
This sequential update of the two replicas and the subsequent 
calculation of $q$ is repeated again and again to measure $P(q)$. 
The simulation is performed until the average of the MC steps of the two replicas 
becomes larger than $5\times 10^7$. The first $1.25\times 10^7$ MC steps are for the equilibration. 
$P(q)$ is calculated in the subsequent $3.75\times 10^7$ MC steps.

{\it Dynamical rules.---}
Because the ergodicity is the possibility for a system to explore the phase space, 
it clearly depends on dynamical rules on which the system explores the phase space. 
As mentioned above, we update the particle configuration with the LRMC method. 
In this method, the particle configuration is changed one by one by either inserting or 
deleting a particle stochastically~\cite{LRMC_DynamicalRule}. An  insertion or deletion site 
is chosen randomly from all of the sites at which we can insert or delete a particle. 
In the present study, we investigate the ergodicity breaking on 
this dynamical rule restricted to the single-particle update.

\begin{figure}[t]
\begin{center}
\includegraphics[width=0.8\columnwidth]{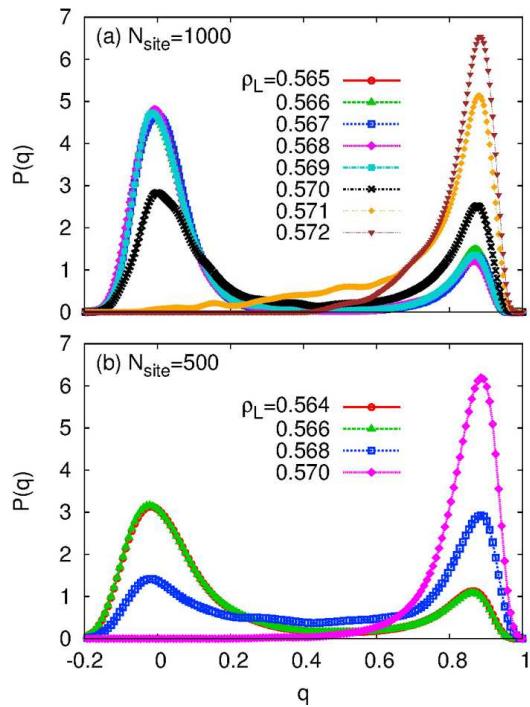}
\end{center}
\caption{(Color online) $\rho_{\rm L}$ dependence of $P(q)$ for (a) $N_{\rm site}=1000$ 
and (b) $N_{\rm site}=500$. The values of $\rho_{\rm U}$ for $N_{\rm site}=1000$ and $N_{\rm site}=500$ 
are $0.573$ and $0.572$, respectively. The average over random graphs is taken over $100$ samples. 
}
\label{fig:PofQ}
\end{figure}

{\it Results.---}
Figure~\ref{fig:PofQ}(a) shows the $\rho_{\rm L}$ dependence of $P(q)$ for $N_{\rm site}=1000$. 
The average over random graphs is taken over $100$ samples.  
$\rho_{\rm U}$ is fixed to $0.573$, which is slightly above the static transition density 
$\rho_{\rm s}=0.5725$ evaluated by the cavity method. 
In contrast, $\rho_{\rm L}$ is changed from $0.565$ to $0.572$ by $0.001$. 
Because $N_{\rm site}$ is $1000$, this is the minimal increment of $\rho_{\rm L}$.  
When $\rho_{\rm L}\le 0.569$, $P(q)$ hardly depends on $\rho_{\rm L}$ and 
all of the data collapse into each other. As expected, $P(q)$ has two peaks 
around $0$ and $1$ (see the top right panel in Fig.~\ref{fig:Expected_Behavior}). 
However, the shape of $P(q)$ abruptly changes at $\rho_{\rm L}=0.570$. 
It is surprising that such a distinct change in $P(q)$ is induced by a minimal change 
in $\rho_{\rm L}$. The peak around zero completely disappears when $\rho_{\rm L}=0.571$ 
and $0.572$. We emphasize that these changes in $P(q)$ are caused solely by the ergodicity breaking 
because $P(q)$ is always measured at the fixed upper occupation density $\rho_{\rm U}=0.573$. 
These observations strongly support that the model exhibits the ergodicity breaking at around 
$\rho_{\rm d}=0.571$, consistent with the previous result by the cavity method.

We next turn to the result of $N_{\rm site}=500$ shown in Fig.~\ref{fig:PofQ}(b). 
In the simulation, $\rho_{\rm L}$ is changed from $0.564$ to $0.570$ by $0.002$, 
which is the minimal increment for $N_{\rm site}=500$. $\rho_{\rm U}$ is fixed to $0.572$. 
We see that $P(q)$ shows a qualitatively similar $\rho_{\rm L}$ dependence to that for $N_{\rm site}=1000$. 
However, there are several quantitative difference between them: Firstly, an abrupt 
change in $P(q)$ occurs at a lower density $\rho_{\rm L}=0.568$. Secondly, $P(q)$ 
has a peak around one even in the ergodic case ($\rho_{\rm L}\le 0.566$) 
despite that $\rho_{\rm U}$ is slightly lower than $\rho_{\rm s}$. We attribute 
these qualitative differences to a finite-size effect. In fact, we have found that the average of 
maximum densities measured in simulations {\it decreases} with decreasing 
$N_{\rm site}$~\cite{SasakiHukushima13}. 

\begin{figure}[t]
\begin{center}
\includegraphics[width=0.8\columnwidth]{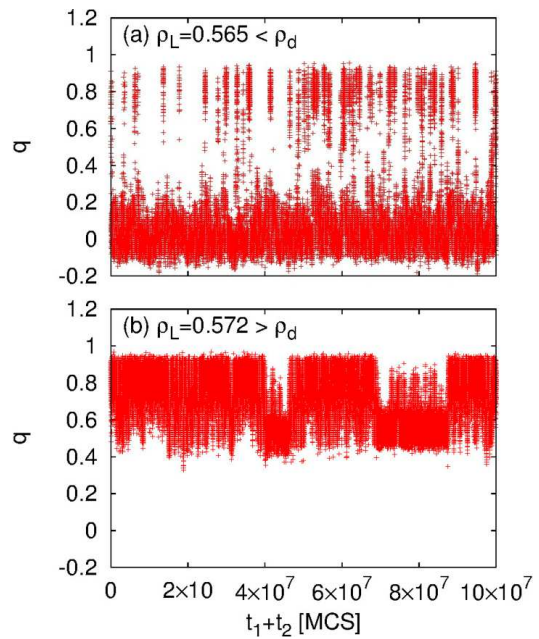}
\end{center}
\caption{(Color online) The overlap $q$ measured at $\rho_{\rm U}=0.573$ 
in one sample is plotted as a function of the sum of the elapsed times 
of the two replicas. The number of sites $N_{\rm site}$ is $1000$. 
The values of $\rho_{\rm L}$ are (a) $0.565$ and (b) $0.572$, respectively. 
A common sample is used for the two cases. 
}
\label{fig:PQvsTime}
\end{figure}

Figure~\ref{fig:PQvsTime} shows how the overlap $q$ changes with MC time 
in the measurement of $P(q)$. In the figure, the overlap $q$ measured in one sample 
is plotted as a function of the sum of the elapsed times of the two replicas. 
The number of sites $N_{\rm site}$ is $1000$. 
The values of $\rho_{\rm L}$ are (a) $0.565$ and (b) $0.572$, respectively. 
A common sample is used for the two cases. 
When (a)~$\rho_{\rm L}=0.565<\rho_{\rm d}$, 
$q$ quickly drops to zero at the beginning and it repeats intermittent increases 
from zero to one. Therefore, $P(q)$ has two peaks in this ergodic case. 
On the contrary, $q$ never drops to zero when (b)~$\rho_{\rm L}=0.572>\rho_{\rm d}$.

{\it Discussion and conclusions.---}
In the present study, we have examined the ergodicity breaking of the BM lattice glass model 
on a regular random graph, and obtained results which nicely agree with those of the cavity method. 
However, we consider that this agreement is not trivial from the following two points of view: 
(i) As mentioned above, the cavity method is based on several non-trivial assumptions. 
(ii) The details of the ergodicity breaking such as the transition density should in principle depend 
on dynamical rules, while they are not specified in the cavity method. 
It is an intriguing issue in the future to unveil how this agreement between our numerical results 
and the analyses based on the cavity method comes from. To this end, it may be helpful to investigate 
how the details of the ergodicity breaking depend on dynamical rules by using the present method. 

In conclusion, we have numerically investigated the ergodicity breaking in the BM lattice glass model 
on a regular random graph. As a result, we have obtained a strong evidence that an ergodicity 
breaking occurs around the dynamical transition density $\rho_{\rm d}=0.5708$ predicted 
by the cavity method. We can easily apply the present numerical method 
to lattice glass models in finite dimensions. We hope that the present study will stimulate 
further research on the ergodicity breaking in glassy systems. 

This work is supported by Grant-in-Aid for Scientific Research Program (No. 22340109, 25400387, 
25610102, and 25120010) from the MEXT in Japan, and by the JSPS Core-to-Core program 
gNon-equilibrium dynamics of soft-matter and informationh.


\begin{thebibliography}{99} 

\bibitem{KirkpatrickThirumalai87}
T.~R.~Kirkpatrick and D.~Thirumalai, Phys. Rev. Lett. {\bf 58}, 2091 (1987).

\bibitem{KirkpatrickWolynes87}
T.~R.~Kirkpatrick and P.~G.~Wolynes, Phys. Rev. B {\bf 36}, 8552 (1987).

\bibitem{KirkpatrickThirumalaiWolynes89}
T.~R.~Kirkpatrick, D.~Thirumalai, and P.~G.~Wolynes, Phys. Rev. A {\bf 40}, 1045 (1989).

\bibitem{MezardParisi01}
M. M{\'e}zard and G. Parisi, Eur. Phys. J. B {\bf 20}, 217 (2001). 

\bibitem{MezardMontanariBook}
M. M{\'e}zard and A. Montanari, {\it Information, Physics, and Computation} 
(Oxford University Press, Oxford, 2009).

\bibitem{BiroliMezard01}
G.~Biroli and M.~M{\'e}zard, Phys. Rev. Lett. {\bf 88}, 025501 (2001). 

\bibitem{Rivoire04}
O.~Rivoire, G.~Biroli, O.~C.~Martin, and M.~M{\'e}zard, Eur. Phys. J. B {\bf 37}, 55 (2004). 

\bibitem{Krzakala08}
F.~Krzakala, M.~Tarzia and L.~Zdeborov{\'a}, Phys. Rev. Lett. {\bf 101}, 165702 (2008).

\bibitem{SasakiHukushima13}
M. Sasaki and K. Hukushima, J. Phys. Soc. Jpn. {\bf 82}, 094003 (2013).

\bibitem{WangLandau01A}
F. Wang and D. P. Landau, Phys. Rev. Lett. {\bf 86}, 2050 (2001). 

\bibitem{WangLandau01B}
F. Wang and D. P. Landau, Phys. Rev. E {\bf 64}, 056101 (2001). 

\bibitem{BortzKalosLebowitz75}
A. B. Bortz, M. H. Kalos, and J. L. Lebowitz, J. Comput. Phys. {\bf 17}, 10 (1975).

\bibitem{LRMC_DynamicalRule}
A particle-hole exchange update in the LRMC method~\cite{SasakiHukushima13} also satisfies 
this {\it one-by-one} rule because the exchange is realized by successively performing a deletion 
and a subsequent insertion of a particle. Furthermore, the restriction on the occupation density 
is imposed not only the initial and terminal states but also the transient state just after the deletion. 

\end{thebibliography}
\end{document}